\documentclass[twocolumn]{autart}    

\usepackage{graphicx}          

\usepackage{amsmath,amssymb,amsfonts}
\usepackage{subfigure}
\usepackage{graphicx}
\graphicspath{{./IMG/}}
\usepackage{epstopdf}
\usepackage{cite}
\usepackage{hyperref}
\usepackage{epstopdf,color}
\usepackage{textcomp}
\usepackage{enumerate}   
\usepackage[numbers]{natbib}

\usepackage{array}
\usepackage{wrapfig}
\usepackage{comment}
\usepackage{amsmath}
\usepackage{etoolbox}
\usepackage{algorithm,algorithmic}
\usepackage{mathrsfs}
\AtBeginEnvironment{algorithm2e}{\algoequations}
\AtEndEnvironment{algorithm2e}{\restoreequations}
\newcounter{algosavedequation}
\newcommand{\algoequations}{%
  \setcounter{algosavedequation}{\value{equation}+1}%
  \setcounter{equation}{0}%
  \renewcommand{\theequation}{\arabic{algosavedequation}\alph{equation}}
}
\usepackage{arydshln}
\newcommand{\restoreequations}{%
  \setcounter{equation}{\value{algosavedequation}}%
}

\allowdisplaybreaks[4]

\newcommand{\longthmtitle}[1]{\mbox{}\emph{(\textbf{#1}):}}

\newtheorem{theorem}{Theorem}[section]
\newtheorem{definition}{Definition}
\newtheorem{assumption}{Assumption}
\newtheorem{lemma}[theorem]{Lemma}
\newtheorem{remark}[theorem]{Remark}
\newtheorem{example}[theorem]{Example}
\newtheorem{corollary}[theorem]{Corollary}
\newtheorem{proposition}[theorem]{Proposition}

\newcommand{\real}{\mathbb{R}} 
\newcommand*{\QEDBL}{\hfill\ensuremath{\blacksquare}}
\newcommand\oprocendsymbol{\hbox{$\square$}}
\newcommand\oprocend{\relax\ifmmode\else\unskip\hfill%
\fi\oprocendsymbol}

\DeclareMathAlphabet{\mymathbb}{U}{BOONDOX-ds}{m}{n}

\newcommand{\zero}{\mymathbb{0}} 

\newcommand{\bx}{{\mathbf{x}}}
\newcommand{\by}{{\mathbf{y}}}
\newcommand{\bu}{{\mathbf{u}}}
\newcommand{\bp}{{\mathbf{p}}}

\newcommand{\bde}{{\boldsymbol{\delta}}}

\newcommand{\Kc}{{\mathcal{K}}}
\newcommand{\Ec}{{\mathcal{E}}}

\newcommand{\Sc}{{\mathcal{S}}}
\newcommand{\Ic}{{\mathcal{I}}}

\newcommand{\setdef}[2]{\{#1 \; : \; #2\}}
\newcommand{\norm}[1]{\Vert #1 \Vert}

\begin{document}

\begin{frontmatter}




\title{Independence of Closed-Loop Equilibria and Stability from the Choice of Control Barrier Function for a Given Safe Set}

\thanks[footnoteinfo]{This work was supported by the AFOSR Award FA9550-23-1-0740. Corresponding author: Yiting Chen. During the preparation of this work, P. Mestres was affiliated with UC San Diego.}

\author[Boston]{Yiting Chen}\ead{yich4684@bu.edu},    
\author[Caltech]{Pol Mestres}\ead{mestres@caltech.edu}, 
\author[SanDiego]{Jorge Cortes}\ead{cortes@ucsd.edu},  
\author[Boston]{Emiliano Dall'Anese}\ead{edallane@bu.edu}
\address[Boston]{Department of Electrical and Computer Engineering, Boston University} 
\address[Caltech]{Department of Mechanical and Civil Engineering, California Institute of Technology} 
\address[SanDiego]{Department of Mechanical and Aerospace Engineering, University of
California, San Diego}

\begin{keyword}               Optimization-based controllers, control barrier functions, safety filters.
\end{keyword}                             

\begin{abstract}
Control barrier functions (CBFs) play a critical role in the design of safe optimization-based controllers for control-affine systems. Given a CBF associated with a 
given, predefined ``safe'' set, the typical approach consists in embedding CBF-based constraints into the optimization problem  defining the control law to enforce forward invariance of the safe set. While this approach effectively guarantees safety for a given CBF, the CBF-based control law can introduce undesirable equilibrium points (i.e., points that are not equilibria of the original system).
 Given that there exist many different CBFs associated with a given fixed safe set, open questions remain  on how the choice of CBF influences the number and locations of undesirable equilibria and, in general, the dynamics of the  closed-loop system. This paper investigates how the choice of CBF impacts  the dynamics of the closed-loop system and shows that: (i) The choice of CBF does not affect the number, location, and (local) stability properties of the equilibria in the interior of the safe set;   (ii)  undesirable equilibria only appear on the boundary of the safe set; and, (iii) the number and location of undesirable equilibria for the closed-loop system do not depend of the choice of the CBF. Additionally, for the well-established \textit{safety filters}, we show that the stability properties of equilibria of the closed-loop system are independent of the choice of the CBF and of the associated extended class-$\Kc$ function,  provided that the CBFs are chosen from the same equivalence class.

\end{abstract}

\end{frontmatter}

\section{Introduction}

Modern control systems for applications ranging from autonomous driving and robotics, to critical infrastructures such as power grids, require the system to satisfy a set of ``safe'' operational constraints. Control barrier functions (CBFs) have emerged as a popular and powerful framework to design controllers that ensure forward-invariance of a given set of states termed as safe~\cite{ADA-SC-ME-GN-KS-PT:19,WX-CGC-CB:23}. CBFs can be used to define affine constraints on the control input. These can then be embedded as constraints of quadratic programs (QPs), termed \emph{safety filters}, where a nominal controller is minimally modified to satisfy the CBF constraint, or in conjunction with control Lyapunov functions (CLFs) in order to guarantee safety and  stability. The ability of CBF-based controllers in ensuring  safety is well investigated in the literature.

However, for any given safe set, there exist multiple CBFs associated with it, and
there is still a limited understanding of how the choice of CBF influences the behavior of the control-affine
system under the CBF-based control law. 
This paper seeks contributions in this direction by investigating the degree to which the choice of  CBF affects the emergence of undesirable equilibria (i.e., spurious equilibria not present in the nominal system and introduced by the CBF-based design) and the local behavior of the closed-loop system.

\subsubsection*{Literature Review}
We rely on the body of work on CBFs
~\cite{PW-FA:07,ADA-SC-ME-GN-KS-PT:19,WX-CGC-CB:23,MK:24-tac}, which are a well-established tool
for rendering a given set forward invariant. A celebrated feature of CBF-based controllers is that they avoid the complex task of computing the system's reachable set, and can be computed efficiently for a variety of control systems.
For example, given a nominal controller with desirable stability or optimality properties, CBFs can act on top of it to ensure safety. This technique is often referred to as \emph{safety filters}~\cite{LW-ADA-ME:17,YC-PM-EDA-JC:24-cdc,WSC-DVD:22-tac}. If the system is control-affine, the controller can be computed by solving a Quadratic Program (QP) at every point of the state space. CBFs have also been combined with CLFs~\cite{EDS:98} in order to design controllers with provable forward invariance and asymptotic stability guarantees to the origin.
These controllers  can also be computed through a QP for control-affine systems and are referred to as CLF-CBF QP controllers. Several works
have thoroughly studied the dynamical properties of the closed-loop systems resulting from CBF-based control designs, establishing conditions under which the origin is asymptotically stable~\cite{WSC-DVD:22-tac,mestres2022optimization}; undesirable equilibria emerge~\cite{MFR-APA-PT:21,XT-DVD:24,YY-SK-BG-NA:23}, and the stability properties of such undesirable equilibria~\cite{YC-PM-EDA-JC:24-cdc,PM-YC-EDA-JC:25-jnls}.

However, there exist many different CBFs associated with any given safe set, and the study of how the properties of CBF-based controllers depend on the choice of CBF remains largely unexplored. 
In fact, all of the aforementioned works presuppose a fixed CBF throughout the analysis.

A key open research question is whether the stability properties of the desirable equilibria and the existence and stability properties of undesirable equilibria are dependent or not on the choice of CBF. 
A notable exception is~\cite[Lemma IV.3]{MA-NA-JC:23-tac}, which shows that the points of discontinuity of CBF-based safety filters are independent of the choice of CBF. This question is of paramount importance for various reasons. If different CBFs induce different undesirable equilibria, one might seek to find the CBF that \textit{minimizes} the number of such undesirable equilibria, for instance. Similarly, if the stability properties of the different equilibria depend on the choice of CBF, one might also seek to find a CBF that makes the desirable equilibria stable (possibly with regions of attraction as large as possible) and the undesirable equilibria unstable.

\begin{figure}[t!]
  \centering 
{\includegraphics[width=0.48\textwidth]{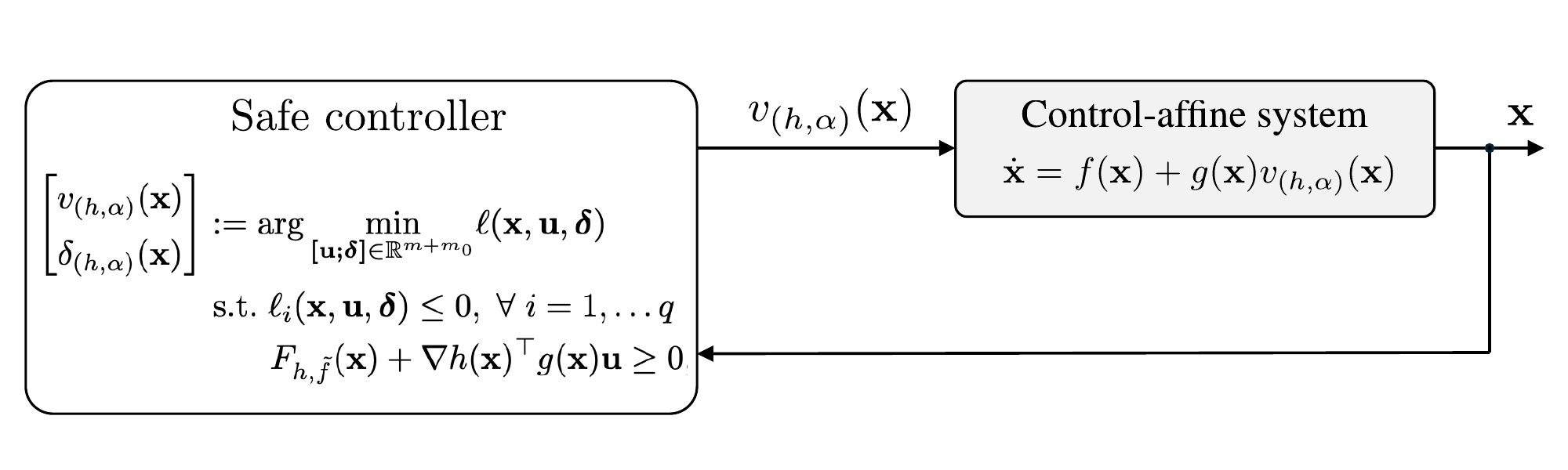}} 
   \vspace{-.4cm}
  \caption{Closed-loop system that is the subject of the paper. We consider a general formulation for the safe optimization-based controllers that subsumes existing safety filters and CLF-CBF QP approaches. }
  \label{fig:F_systems}
\end{figure}

\subsubsection*{Statement of Contributions}

Our contributions are as follows.

$\bullet$ 
We consider a control-affine system with a  predefined and fixed safe set, along with a
general safe optimization-based controller as shown in Figure~\ref{fig:F_systems} (notation will be introduced shortly). The safe controller is ``generalized'', in the sense that it subsumes existing optimization-based approaches, including the CLF-CBF QP and safety filters. Under a general set of assumptions and for a wide range of CBF-based controllers, we  show that: 

\begin{enumerate}[(i)]
    \item The number and location of undesirable equilibria of the closed-loop system are independent of the choice of
CBF for the given safe set. 
    
    \item  
    The number, location, and stability properties of the desired equilibria are independent of the choice of CBF for the given safe set.

        \item 
        The undesirable equilibria (if they exist) only appear on the boundary of the safe set, and the dynamics of the closed-loop system on the boundary of the safe set depends on the geometry of safe set, rather than the  choice of CBF. Therefore, the trajectories that remain in the boundary of the safe set are also independent of the choice of the CBF.

\end{enumerate}

$\bullet$ For safety filters, we also show that the stability properties of the undesirable equilibria are independent of the choice of CBF. Although the dynamical properties of closed-loop systems obtained from this design are well studied in the literature for a fixed CBF, the results presented here are novel.
Furthermore, we provide explicit expressions for the Jacobian of the closed-loop system evaluated at undesirable equilibria, which show how the dynamics, the CBF, and the different parameters used in the control design affect the stability properties of undesirable equilibria. 

$\bullet$ Finally, we note that some intermediate results in Section~\ref{sec:preliminaries} are of independent interest, and they characterize the relationship between the gradients and Hessian matrices of two CBFs for a given safe set when evaluated at its boundary.

\section{Preliminaries}\label{sec:preliminaries}
In this section, we outline the notation used throughout the paper and relevant definitions, and provide some new, intermediate technical results that will be used in the paper.  
\subsection{Notation} 
We denote by $\mathbb{Z}_{+}$, $\real$ and $\real_{\geq0}$ the set of positive integers, real, and nonnegative real numbers, respectively.  
 We write $\text{int}(\mathcal{S}), \partial\mathcal{S}$ and $\Sc^c$ for the interior, boundary, and complement of the set $\mathcal{S}$, respectively. 
Formally, $
\operatorname{int}(\Sc) = \{\, \bx \in \Sc : \exists \,\varepsilon > 0 \ \text{such that}\ B(\bx,\varepsilon) \subseteq \Sc \,\},
$
 $
\partial \Sc = \{\, \bx \in \Sc : \forall \varepsilon > 0,\ (B(\bx,\varepsilon)\cap \Sc \neq \varnothing) \ \text{and}\ (B(\bx,\varepsilon)\cap \Sc^c \neq \varnothing) \,\},$  where $B(\bx,\varepsilon)$ is the open ball centered at $\bx$ with radius $\varepsilon$.
Throughout the paper, boldface symbols denote vectors of finite dimension and non boldface symbols denote scalar values.
We let $\textbf{0}_n$ be the $n$-dimensional zero vector.
Given $\bx\in\real^{n}$, $\norm{\bx}$ denotes its Euclidean norm; for a positive definite matrix $G\in\real^{n\times n}$, we define $\norm{\bu}_G = \sqrt{\bu^\top G \bu}$.
A function $\beta:\real\to\real$ is of extended class $\mathcal{K}_{\infty}$ if $\beta(0)=0$, it is strictly increasing, and 
$\lim\limits_{t\to\pm\infty}\beta(t)=\pm\infty$.
A function $V:\real^{n}\to\real$ is
positive definite if $V(\textbf{0}_n)=\textbf{0}_n$ and $V(\bx)>0$ for $\bx \neq \textbf{0}_n$.
Let $a:\real^n\times\real^m\to\real$, $b:\real^n\times\real^m\to\real^q$, and consider the optimization problem
\begin{subequations}\label{eq:optim-based-controller}
  \begin{align}
    &   \underset{\bu\in\real^m}{\text{argmin}} \ \
    a(\bx,\bu)
  \\
  & \qquad \text{s.t.} \ b(\bx,\bu) \leq \textbf{0}_q \label{eq:optim-based-controller-const}
\end{align}  
\end{subequations}
for a given $\bx\in\real^n$, where the inequality in~\eqref{eq:optim-based-controller-const} is entry-wise. Then, given $\bu\in\real^m$, we let $\Ic(\bx,\bu)$ be
the set of active constraints of~\eqref{eq:optim-based-controller} at $(\bx,\bu)$, i.e.,
$\Ic(\bx,\bu):=\setdef{i\in\{1,\hdots q\} }{{b}_i(\bx,\bu) = 0}$.
Slater's condition holds at $\bx$ for Problem~\eqref{eq:optim-based-controller} if there exists $\hat{\bu}\in\real^m$ such that $b_i(\bx,\hat{\bu}) < 0$ for all $i\in\{1,\hdots, q \}$.
The Linear Independence Constraint Qualification (LICQ) holds at $(\bx,\bu)\in\real^n\times\real^m$ for Problem~\eqref{eq:optim-based-controller} if the vectors $\{ \nabla_{\bu} {b}_i(\bx,\bu), i\in\Ic(\bx,\bu) \}$ are linearly independent. Given $\bx\in\real^n$, a point $(\bu_{\bx}, \lambda_{\bx})\in\real^m\times\real^q$ is a Karush-Kuhn-Tucker (KKT) point of~\eqref{eq:optim-based-controller} at $\bx$ if it satisfies the following conditions, which we refer to as KKT equations:
\begin{subequations}
\begin{align}
    &\nabla_{\bu} a(\bx,\bu_{\bx}) + \frac{\partial b}{\partial \bu}(\bx,\bu_{\bx})^\top \lambda_{\bx} = \textbf{0}_m, \\
    &b(\bx,\bu_{\bx})\leq\textbf{0}_q, \ \lambda_{\bx} \geq \textbf{0}_q, \\
    &\lambda_{\bx}^\top b(\bx,\bu_{\bx}) = 0.~\label{eq:complementary-slackness}
\end{align}
\label{eq:kkt-equations-optim-based-controller}
\end{subequations}
We refer to~\eqref{eq:complementary-slackness} as the \textit{complementary slackness} condition.

\subsection{Control Barrier Functions}

Consider a  nonlinear control-affine dynamical system
\begin{align}\label{eq:control-affine-sys}
  \dot{\bx}=f(\bx)+g(\bx)\bu,
\end{align}
where $f:\real^{n}\to\real^{n}$ and $g:\real^{n}\to\real^{n\times m}$
are locally Lipschitz functions, $\bx\in\real^{n}$ is the state, and
$\bu\in\real^{m}$ is the input. We let $\mathcal{S}\subset\real^n$ be the safe set.

\vspace{.1cm}

\begin{definition}\longthmtitle{Control Barrier Function}\label{def:cbf}
    Let $h:\real^n\to\real$ be a continuously differentiable function such that $\mathcal{S}=\{\bx\in\real^n:h(\bx)\geq 0\}$.
    The function $h$ is a Control Barrier
    Function \textbf{(CBF)} of the set $\mathcal{S}$ for the system~\eqref{eq:control-affine-sys} if there exists an extended class
  $\mathcal{K}_{\infty}$ function $\alpha$ such that, for each $\bx\in\mathcal{S}$, there exists a control $\bu\in\real^m$ satisfying
  $\nabla h(\bx)^\top (f(\bx)+g(\bx)\bu) + \alpha(h(\bx)) \geq 0$. 
   Moreover, if for every $\bx\in\mathcal{S}$, there exists $\bu$ such that the inequality is satisfied strictly, then we call $h$ strict CBF and the pair $(h,\alpha)$ strict CBF pair.
  \hfill $\Box$
\end{definition}

Given a safe set $\mathcal{S}$,
if $h:\real^n\to\real$ is a CBF of $\mathcal{S}$ and the pair $(h,\alpha)$ satisfies Definition~\ref{def:cbf}, then for any $a_1\geq 1$, $a_2\geq a_1$, the pair $(a_1 h,a_2\alpha)$ satisfies Definition~\ref{def:cbf} too. 

Indeed, let $\bu_{\bx}$ be such that $\nabla h(\bx)^\top (f(\bx)+g(\bx)\bu_{\bx}) + \alpha(h(\bx)) \geq 0$.
Then, we have
\begin{align*}
    &a_1 \nabla h(\bx)^\top (f(\bx)+g(\bx)\bu_{\bx}) + a_2 \alpha (a_1 h(\bx)) \geq \\
    &a_1 \Big( \nabla h(\bx)^\top (f(\bx)+g(\bx)\bu_{\bx}) + \alpha(a_1 h(\bx)) \Big) \geq \\
    &a_1 \Big( \alpha(a_1 h(\bx)) - \alpha(h(\bx)) \Big) \geq 0,
\end{align*}
where in the last inequality we have used the fact that $a_1 \geq 1$ and $\alpha$ is increasing.
Hence, $\bu_{\bx}$ also satisfies the CBF condition for the pair $(a_1 h, a_2 \alpha)$.
Therefore, if $h$ is a CBF of $\Sc$, then
there are multiple (in fact, infinitely many) pairs satisfying Definition~\ref{def:cbf}.

The next result establishes the regularity properties of strict CBFs.

\begin{lemma}\longthmtitle{Regularity of Strict CBF}
\label{lem: boundary matching}
Let $h$ be any strict CBF of $\mathcal{S}$, then: (i) For any $\bx\in\Sc$, it holds that $\nabla h(\bx)\neq \mathbf{0}_n$ if $h(\bx)=0$; (ii) $\partial\Sc=\setdef{\bx\in\real^n}{h(\bx)=0}$; (iii) $\operatorname{Int}(\mathcal{S})=\setdef{\bx\in\real^n}{h(\bx)>0}$.
\end{lemma}
\vspace{-0.5cm}
\begin{pf}
 To show (i), suppose there exists a point $\bp$ such that $\nabla h(\bp)=\mathbf{0}_n$ and $h(\bp)=0$. Then the  CBF constraint can not  hold strictly at $\bp$, which is a contradiction.   Next, we prove (ii).
By continuity of $h$, it follows that $\{\bx:h(\bx)>0\}\subseteq\operatorname{Int}(\mathcal{S})$. Hence we have $\partial\mathcal{S}\subseteq \{\bx:h(\bx)=0\}$. Then it suffices to show that $ \{\bx:h(\bx)=0\}\cap \operatorname{Int}(\mathcal{S})=\varnothing$. Suppose that there exists $\bp\in\operatorname{Int}(\mathcal{S})$ such that $h(\bp)=0$. We claim that $\nabla h(\bp)=0$. Otherwise, since $\bp\in\operatorname{Int}(\mathcal{S})$, we can find a small enough $\epsilon>0$ such that $h(\bp-\epsilon \nabla h(\bp))<h(\bp)=0$ and $\bp-\epsilon \nabla h(\bp)\in \mathcal{S}$. We show (ii) by noting that $\nabla h(\bp)=0$, together with $h(\bp)=0$, contradicts part (i). Finally, (iii) follows directly from (ii) and $\mathcal{S}=\{\bx\in\real^n:h(\bx)\geq 0\}$.
\hfill $\Box$
\end{pf}
\vspace{-0.5cm}
 Next, given two CBFs of $\mathcal{S}$, the following result presents the
relationship between their gradients evaluated at $\partial \mathcal{S}$.

\begin{lemma}\longthmtitle{Relation between Gradients of CBFs}
\label{lem: the existence of positive scalar function}
    Let $h_1,h_2:\real^n\to\real$ be two strict CBFs of $\Sc$.
    Then $\nabla h_2(\bx)=\zeta(\bx) \nabla h_1(\bx)$, with $\zeta :\partial \mathcal{S}\mapsto \mathbb{R}_{>0}$ a function that is unique.
\end{lemma}
\begin{pf}
    We follow a similar argument to the one in~\cite[Lemma IV.3]{MA-NA-JC:23-tac}. 
    By Lemma~\ref{lem: boundary matching},~\cite[Theorem 5.1]{MS:95} and the fact that
     $\nabla h_1(\bx) \neq \mathbf{0}_n$ and $\nabla h_2(\bx) \neq \mathbf{0}_n$ for all $\bx \in \partial\mathcal{S}$, the sets $\setdef{ \bx\in\real^n }{ h_1(\bx)=0 }$ and $\setdef{ \bx\in\real^n }{ h_2(\bx)=0 }$ define the same differentiable manifold $\partial\Sc$ of dimension $n-1$ embedded in $\real^n$. By~\cite[Theorem 3.15]{NB:23}, the tangent space of $\partial\Sc$ at a point $\bx\in\partial\Sc$ is given by $T_{\bx} = \text{ker}(\nabla h_1(\bx) ) = \text{ker}(\nabla h_2(\bx) )$. This implies that $\nabla h_1(\bx)$ and $\nabla h_2(\bx)$ are parallel. Moreover, since $h_1$ and $h_2$ have the same $0$-superlevel set, it follows that $\nabla h_2(\bx)^\top \nabla h_1(\bx) > 0$.
    This implies that there exists a unique function $\zeta:\partial\Sc\to\real_{>0}$ satisfying $\nabla h_2(\bx) = \zeta(\bx) \nabla h_1(\bx)$ for all $\bx\in\partial\Sc$. \hfill $\Box$
\end{pf}

Given the result in Lemma \ref{lem: the existence of positive scalar function}, for two CBFs $h_1$ and $h_2$ of the set $\mathcal{S}$, we let $ \zeta_{(h_1,h_2)}:\partial\mathcal{S}\mapsto \mathbb{R}_{>0}$ be such that $\nabla h_2(\bx) =\zeta_{(h_1,h_2)}(\bx) \nabla h_1(\bx)$ for each $\bx\in\partial\Sc$.
Next, we study the relationship between the Hessians of two CBFs. To motivate it, we consider the following example.

\begin{example}\longthmtitle{Relation between Hessians of CBFs}
{\rm
Consider a ball obstacle with center $\bx_c$ and radius~$r_0$. Consider the CBFs $h_1(\bx)=\|\bx-\bx_c\|_2^2-r_0^2$ and $h_2(\bx)=(\|\bx-\bx_c\|_2^2-r_0^2)r(\bx)$, with $r(\bx):=\|\bx\|_2^2+1$. Then, for any $\bx\in\partial\Sc$, $\nabla h_1(\bx)=2(\bx-\bx_c)$ and $\nabla h_2(\bx)=r(\bx)\nabla h_1(\bx)+h_1(\bx)\nabla r(\bx)=r(\bx)\nabla h_1(\bx)$ due to $h_1(\bx)=0$. Hence, $ \zeta_{(h_1,h_2)}(\bx)=r(\bx)$ in this case. If we further evaluate the Hessian matrix at $\bx\in\partial \Sc$, we get that $H_{h_1}(\bx)=2\textbf{I}_n$ and 
\begin{align*}
&H_{h_2}(\bx)  =  \nabla h_1(\bx) \nabla r(\bx)^\top \! + \nabla r(\bx) \nabla h_1(\bx)^\top \\
&~~~~~~~~~~~~~~~~+r (\bx) H_{h_1}(\bx)+h_1 (\bx) H_{r}(\bx)\\
&=  \nabla h_1(\bx) \nabla r(\bx)^\top+\nabla r(\bx) \nabla h_1(\bx)^\top  +r (\bx) H_{h_1}(\bx).
\end{align*}
Therefore, the difference between the Hessian of $h_1$ multiplied by $ \zeta_{(h_1,h_2)}(\bx)$ and the Hessian of $h_2$ evaluated at $\partial\Sc$ is equal to the sum of a rank-one matrix and its transpose.
\hfill $\Box$}
\end{example}

Inspired by the above example, we define a relation between two CBFs. Formally, given any two pairs $(h_1,\alpha_1)$ and $(h_2,\alpha_2)$ satisfying Definition~\ref{def:cbf}, we use the notation $h_1\stackrel{\text{H}}{\sim}h_2$ if there exists $\zeta(\bx):\partial \mathcal{S}\mapsto \mathbb{R}_{>0}$ and  
$\tilde{\zeta}(\bx):\partial \mathcal{S}\mapsto \mathbb{R}^{n}$
such that for all $\bx \in \partial\mathcal{S}$,   $\nabla h_2(\bx)=\zeta(\bx)\nabla h_1(\bx)$ and 
\[ H_{h_2}(\bx)=  \nabla h_1(\bx) \tilde{\zeta}(\bx)^\top + \tilde{\zeta}(\bx) \nabla h_1(\bx)^\top +\zeta(\bx) H_{h_1}(\bx). \]
We have the following general result. 
\vspace{-0.3cm}
\begin{proposition}\longthmtitle{Equivalence Relation}
\label{prop: generalized equivialnece relation of Hessian}
    $ \stackrel{\text{H}}{\sim} $ is an equivalence relation. 
\end{proposition}
\vspace{-0.6cm}
\begin{pf}
We need to show that  ``$ \stackrel{\text{H}}{\sim} $'' is (a) reflexive: $h\stackrel{\text{H}}{\sim} h$ for any $h$; (b) symmetric: $h_2 \stackrel{\text{H}}{\sim} h_1$ if $h_1 \stackrel{\text{H}}{\sim} h_2$;  and (c) transitive: $h_1\stackrel{\text{H}}{\sim}h_3$ if $h_1\stackrel{\text{H}}{\sim} h_2$ and $h_2 \stackrel{\text{H}}{\sim}h_3$.

\emph{(a) Reflexivity.} Taking $\zeta \equiv 1$ and $\tilde{\zeta} \equiv 0$, it follows that $h\stackrel{\text{H}}{\sim} h$.

\emph{(b) Symmetry.} Suppose that $\nabla h_2(\bx)=\zeta(\bx)\nabla h_1(\bx)$ and $H_{h_2}(\bx)=  \nabla h_1(\bx) \tilde{\zeta}(\bx)^\top + \tilde{\zeta}(\bx) \nabla h_1(\bx)^\top +\zeta(\bx) H_{h_1}(\bx)$. It follows that $\nabla h_1(\bx)=\frac{1}{\zeta(\bx)}\nabla h_2(\bx)$ and
\begin{align*}
    H_{h_1}= \frac{1}{\zeta} H_{h_2}  -  \nabla h_2  \frac{\zeta_{(h_2,h_1)}}{\zeta} \tilde{\zeta}^\top - \frac{\zeta_{(h_2,h_1)}}{\zeta}\tilde{\zeta} \nabla h_2^\top , 
\end{align*}
where $\zeta_{(h_2,h_1)}$ satisfying $\nabla h_1(\bx)=\zeta_{(h_2,h_1)}(\bx)\nabla h_2(\bx)$ by Lemma~\ref{lem: the existence of positive scalar function}.

\emph{(c) Transitivity.} Suppose that 
$\nabla h_2(\bx)=\zeta_1(\bx)\nabla h_1(\bx)$, $     H_{h_2}(\bx)=  \nabla h_1(\bx) \tilde{\zeta}_1(\bx)^\top + \tilde{\zeta}_1(\bx) \nabla h_1(\bx)^\top +\zeta_1(\bx) H_{h_1}(\bx)$, $\nabla h_3(\bx)=\zeta_2(\bx)\nabla h_2(\bx)$ , $      H_{h_3}(\bx)=  \nabla h_2(\bx) \tilde{\zeta}_2(\bx)^\top + \tilde{\zeta}_2(\bx) \nabla h_2(\bx)^\top +\zeta_2(\bx) H_{h_2}(\bx)$.
It follows that $\nabla h_3(\bx)=\zeta_2(\bx)\zeta_1(\bx)\nabla h_1(\bx)$ and
\vspace{-0.2cm}
\begin{align*}
      H_{h_3}=&  \nabla h_1 \zeta_{(h_1,h_2)} \tilde{\zeta_2}^\top + \tilde{\zeta}_2 \zeta_{(h_1,h_2)} \nabla h_1^\top \\
      &+\zeta_2 (\nabla h_1 \tilde{\zeta}_1^\top + \tilde{\zeta}_1 \nabla h_1^\top +\zeta_1 H_{h_1})\\
      =& \nabla h_1(\zeta_{(h_1,h_2)} \tilde{\zeta}_2+  \zeta_2\tilde{\zeta}_1 )^\top\\
      &+( \tilde{\zeta}_2 \zeta_{(h_1,h_2)}+ \zeta_2\tilde{\zeta}_1   )\nabla h_1^\top+\zeta_2\zeta_1 H_{h_1}.\qquad \qquad \Box
\end{align*}
\end{pf}
Next, we characterize a class 
of CBFs which are equivalent in the sense of $ \stackrel{\text{H}}{\sim} $.

Given a pair $(h,\alpha)$,
a continuously differentiable extended class $\Kc$ function $\gamma:\real\to\real$, a continuously differentiable positive function $\eta:\real^n\to\real$, and $a,b\in\mathbb{R}_{\geq 0}$ with $a+b>0$, we consider the functional operation $\phi(h)$, 
where $\phi(h):\real^n\to\real$ is given by
$\phi(h)(x)=a \gamma(h)(x)+b \eta(x) h(x)$ for all $x\in\real^n$.
Given a continuously differentiable function $h$, its transform $\phi(h)$ is also a continuously differentiable function with the same domain and the same $0$-superlevel set. 
We should point out that $\phi(h)$ may not be a valid CBF, even if $h$ is.
The following result shows that if $h$ and $\phi(h)$ are CBFs, then they are equivalent.

\begin{proposition}\longthmtitle{Conditions for Equivalence of CBF and its Transform}
\label{Prop: Hessian-2}
    Let $h_1:\real^n\to\real$, $h_2:\real^n\to\real$ be two strict CBFs of $\Sc$. Suppose that there exists $\gamma:\real\to\real$, $\eta:\real^n\to\real$, $a\geq0$, and $b\geq0$ such that:
    \begin{itemize}
        \item $h_2(\bx)= a\gamma(h_1(\bx))+b\eta(\bx)h_1(\bx)$, for all $\bx\in\mathbb{R}^n$;
        \item $\gamma$ is twice continuously differentiable on an open set containing $0$;
        \item $\eta$ is twice continuously differentiable on an open set containing $\partial\mathcal{S}$;
        \item $a+b>0$.
    \end{itemize}
    \vspace{-.3cm}
   Then $h_1\stackrel{\text{H}}{\sim} h_2$.
\end{proposition}
\vspace{-0.4cm}
\begin{pf}
By differentiating $h_2(\bx)= a\gamma(h_1(\bx))+b\eta(\bx)h_1(\bx)$ we obtain
\begin{align*}
    \nabla h_2(\bx) = (a\gamma^{\prime}( h_1(\bx) ) + b\eta(\bx) )\nabla h_1(\bx) + b h_1(\bx) \nabla \eta(\bx) \, .
\end{align*}
By differentiating one more time we obtain the desired expression by taking $\zeta(\bx) = a\gamma^{\prime}(0) + b \eta(\bx)$, $\tilde{\zeta}(\bx) = \frac{1}{2}a \gamma^{\prime\prime}(0) \nabla h_1(\bx) + b \nabla \eta(\bx)$. 
\hfill $\Box$ \end{pf}

By Proposition~\ref{Prop: Hessian-2}, given a CBF $h$, one can define a class of functions 
\begin{align*}
    &\Psi^1(h):=\{a\cdot\gamma(h)+b\cdot\eta \cdot h:~\text{$\forall~a,b,\gamma,\eta$ satisfying}\\
    &~~~~~~~~\text{assumptions in Proposition~\ref{Prop: Hessian-2}}\},\\
    &\Psi^i(h):=\{a\cdot\gamma(\tilde{h})+b\cdot\eta \cdot \tilde{h}:~\text{$\forall~a,b,\gamma,\eta$ satisfying}\\
     &~~~~~\text{assumptions in Proposition~\ref{Prop: Hessian-2}, $\forall~\tilde{h}\in\Psi^{i-1}(h)$}\},
\end{align*}
for any positive integer $i\geq 2$. Here, $\Psi^i(h)$ is the set of function obtained from applying $i$ times on $h$ an operation of the type $\phi$, 
where $a,b,\gamma,\eta$ can be different in each step of the composition. The following result is a consequence of Propositions~\ref{prop: generalized equivialnece relation of Hessian} and \ref{Prop: Hessian-2}.

\begin{corollary}\longthmtitle{Equivalence between the Functions in $\bigcup_{i\in\mathbb{Z}_+} \Psi^i(h)$}\label{corollary: equivialent class of fucntion}
    Let $h$, $h_1$, and $h_2$ be strict CBFs of a safe set $\Sc\subset\real^n$.
    Suppose that $h_1,h_2\in\bigcup_{i\in\mathbb{Z}_+} \Psi^i(h)$. Then $h_1\stackrel{\text{H}}{\sim} h_2$. \hfill $\Box$
\end{corollary}

Corollary~\ref{corollary: equivialent class of fucntion} asserts that, if two CBFs are synthesized from multi-composite operations based on the same CBF $h$, then the two CBFs are equivalent.

\section{Problem Statement}\label{sec:problem-statement}

Consider a  nonlinear control-affine system~\eqref{eq:control-affine-sys} with $f$ and $g$ continuously differentiable subject
to $\bu=v(\bx)$,
\begin{align}\label{eq:unfiltered-system}
    \dot{\bx} = {f}(\bx) + g(\bx)v(\bx),
\end{align}
where $v:\real^n\to\real^m$ is  defined as the unique solution of the following optimization problem 
\begin{align}\label{eq:v-w/o-cbf} 
\begin{bmatrix}
    v(\bx)\\ \delta(\bx)
\end{bmatrix}&=\arg\underset{ \boldsymbol{[\bu; \boldsymbol{\delta}]} \in \mathbb{R}^{m+m_0}}{\min}  \ell(\bx,\bu,\boldsymbol{\delta}) \\
\notag
&\text { s.t. } \ell_i(\bx,\bu,\boldsymbol{\delta} )\leq 0,~\forall i = 1,\ldots q.
\end{align}
Here, $\ell:\real^n\times\real^m\times\real^{m_0}\to\real$ is the objective function and $\ell_i:\real^n\times\real^m\times\real^{m_0}\to\real$ for $i=1,\hdots,q$, $q\in\mathbb{Z}_{+}$ are the constraints (assumptions on objective and constraints will be explained shortly).
Moreover, $\bx$ is the state and $\boldsymbol{\delta}$ is an auxiliary optimization variable. Assume that $\bx^*\in\real^n$ is such that ${f}(\bx^*) + g(\bx^*)v(\bx^*)=0$.  The computation of $v(\bx)$ via~\eqref{eq:v-w/o-cbf}  encompasses various existing optimization-based control methods as special cases.

\begin{remark}\longthmtitle{Special cases of~\eqref{eq:v-w/o-cbf}}
\label{rem:examplesopt}
{\rm
Examples of existing optimization-based controllers that are modeled by~\eqref{eq:v-w/o-cbf} include the following.
\begin{description}
    \item[Pointwise minimum-norm control:]  
    Given a CLF $V$ for system~\eqref{eq:control-affine-sys} with respect to a certain point $\bx^*$, the pointwise minimum-norm controller~\cite[Chapter 4.2]{RAF-PVK:96a} finds the minimum-norm controller satisfying the CLF condition. This can be obtained from~\eqref{eq:v-w/o-cbf} by taking  $\ell(\bx,\bu,\boldsymbol{\delta})=\norm{\bu}^2 + \delta^2$, $q=1$ and $\ell_1(\bx,\bu,\delta) = \nabla V(\bx)^\top (f(\bx)+g(\bx)\bu) + \beta(V(\bx))$. In this case, for the auxiliary variable, the optimal solution is trivially $\delta(\bx) = 0$.  Moreover, if the CLF inequality is infeasible, one may use instead $\ell_1(\bx,\bu,\delta) = \nabla V(\bx)^\top (f(\bx)+g(\bx)\bu) + \beta(V(\bx))-\delta$.  
\item[Stabilization under model uncertainty:]
    If system~\eqref{eq:control-affine-sys} is unknown but the uncertainty can be modeled as a Gaussian Process, and a CLF $V$ for the true system~\eqref{eq:control-affine-sys} is available,~\cite{FC-JJC-BZ-CJT-KS:21-acc} shows that a controller of the form~\eqref{eq:v-w/o-cbf}
    can be used to robustly stabilize the system.
    \item[Input constraints:] 
    Consider the system $\dot{\bx} = f(\bx) + g(\bx)  v(\bx)$ in~\eqref{eq:unfiltered-system}, with $f(\bx) = F(\bx) + g(\bx)k(\bx)$, and 
   $k(\bx)$ is a nominal controller designed based on given control objectives. Suppose that the input produced by $\bu = k(\bx)$ has to be limited to a convex set $\mathcal{U} = \{\bu \in \mathbb{R}^m: c_i^\top \bu \leq d_i, i = 1,\ldots, q\}$, where $c_i \in \mathbb{R}^m$ and  $d_i \in \mathbb{R}$. Input saturation is included in this model. Then, here~\eqref{eq:v-w/o-cbf} can perform the projection of $k(\bx)$ onto $\mathcal{U}$ by setting $\ell(\bx,\bu,\delta)=\norm{\bu - k(\bx)}^2 + \delta^2$ and $\ell_i(\bx,\bu,\delta) = c_i^\top \bu \leq d_i$, $i = 1,\ldots, q$.  
\end{description}
Note also that one can ``deactivate" the controller; i.e., set $v(\bx) = 0$. This can be done by simply setting  $\ell(\bx,\bu,\bde)=\|\bu\|_2^2 + \|\bde\|_2^2$ and $\ell_i\equiv 0$, yielding $v(\bx)= 0$ and $\delta(\bx) = 0$ for any $\bx$ in~\eqref{eq:v-w/o-cbf}. 
    \hfill $\Box$    
    }
\end{remark}

In general, the closed-loop dynamics 
${f}+gv$ might enjoy desirable properties such as stability, optimality, or robustness, but might also be unsafe. We seek to modify the optimization problem defining $v(\bx)$ to achieve safety guarantees. Before presenting our control design,
let $(h,\alpha)$ be a pair satisfying Definition~\ref{def:cbf}. We  make the following assumptions.

\begin{assumption}\longthmtitle{Differentiability and Convexity of Objective Function and Constraints}\label{as:diff-convexity}
The functions $\ell:\real^n\times\real^m\times\real^{m_0}\to\real$ and $\{\ell_i:\real^n\times\real^m\times\real^{m_0}\to\real\}_{i=1}^q$ are such that: 
    \begin{itemize} 
        \item for any fixed $\bx\in\real^n$ and any $1\leq i\leq q$, $\ell_i(\bx,\cdot,\cdot)$ is convex w.r.t $[\bu;\bde]$
        and $\ell(\bx,\cdot,\cdot)$ is strongly convex w.r.t $[\bu;\bde]$.

        \item $\ell$, $\nabla_{[\bu;\bde]} \ell$, $\ell_i$ and $\nabla_{[\bu;\bde]} \ell_i$ ,$i=1,...,q$, are continuously differentiable w.r.t $[\bx;\bu;\bde]$;

        \hfill $\Box$
    \end{itemize}   
\end{assumption}

Next, we consider a modification of the optimization-based controller~\eqref{eq:v-w/o-cbf} by considering CBF-based safety constraints. This leads to  the following optimization problem: 
\begin{align} \label{eq:general-v-problem} 
\begin{bmatrix}
    v_{(h,\alpha)}(\bx) \\ \delta_{(h,\alpha)}(\bx)
\end{bmatrix}&:=\arg\underset{ \boldsymbol{[\bu; \boldsymbol{\delta}]} \in \mathbb{R}^{m+m_0}}{\min} \ell(\bx,\bu,\boldsymbol{\delta})  \\
\notag
& \text { s.t. }  \ell_i(\bx,\bu,\boldsymbol{\delta} )\leq 0,~\forall~ i = 1,\ldots q \\
\notag
&\qquad F_{h,{f}}(\bx)+\nabla h(\bx)^\top g(\bx)\bu  \geq 0,
\end{align}
where $F_{h,{f}}(\bx) = \nabla h(\bx)^\top {f}(\bx) + \alpha(h(\bx)).$ Again, \eqref{eq:v-w/o-cbf} is the version of~\eqref{eq:general-v-problem} without the CBF constraint, which is introduced to achieve safety guarantees. We note that several CBF-based control designs, such as penalty function-based~\cite{mestres2022optimization}, input-constrained~\cite{DRA-DP:21}, and multiple CBF-based controllers~\cite{XT-DVD:22-cdc} can be cast as~\eqref{eq:general-v-problem} with an appropriate choice of the functions $\ell$ and $\{ \ell_i \}_{i=1}^q$.  For instance, CLF-CBF quadratic programs~\cite{ADA-XX-JWG-PT:17,XT-DVD:24} correspond to  $m_0=1$, $q=1$, $\ell_1(\bx,\bu,\boldsymbol{\delta}) = F_{V,{f}}(\bx)+\nabla V(\bx)^\top g(\bx)\bu - \delta$ and  $\ell(\bx,\bu,\boldsymbol{\delta}) = \frac{1}{2}\| \boldsymbol{u}  \|_{G(\bx)}^2 +\frac{1}{2}p\delta^2$, where $V:\real^n\to\real$ is a CLF. Safety filters, which we explicitly deal with in Section~\ref{sec:choice-safety-filters}, are yet another class of designs that can be cast as~\eqref{eq:general-v-problem}.

Next, we make the following assumption on~\eqref{eq:general-v-problem}.

\begin{assumption}\longthmtitle{Slater's Condition}\label{as:constraint-qualif}
    For all $\bx\in\Sc$, problem~\eqref{eq:general-v-problem} satisfies Slater's condition. \hfill $\Box$
\end{assumption}
Note that Assumption~\ref{as:constraint-qualif} implies that problem~\eqref{eq:v-w/o-cbf}  satisfies Slater's condition as well; moreover,  Assumption~\ref{as:constraint-qualif} 
is equivalent to~\eqref{eq:general-v-problem} and~\eqref{eq:v-w/o-cbf} being strictly feasible (i.e, all inequalities being feasible with a strict inequality).
Note further that, for any given  $\bx\in\Sc$, the optimization problems~\eqref{eq:general-v-problem} and~\eqref{eq:v-w/o-cbf} each have a unique solution; this means that $v_{(h,\alpha)}(\bx)$ and $v(\bx)$ are well-defined for all $\bx\in\Sc$. 

Now, consider the  nonlinear system
\begin{align}\label{eq:general-system-2}
    \dot{\bx} = {f}(\bx) + g(\bx)v_{(h,\alpha)}(\bx),
\end{align}
which we refer to as the \textit{filtered system} (because the constraint $F_{h,{f}}(\bx)+\nabla h(\bx)^\top g(\bx)\bu  \geq 0$ is included in the computation of the control input). Correspondingly, we refer to \eqref{eq:unfiltered-system} as the \textit{unfiltered system}. Given any pair $(h,\alpha)$ satisfying Definition~\ref{def:cbf}, the set $\mathcal{S}$ is forward-invariant for the filtered system but may not be forward-invariant for the unfiltered one. Throughout the paper, we refer to an equilibrium of the unfiltered system ~\eqref{eq:unfiltered-system} 
as a \textit{desirable equilibrium} and
an equilibrium of the filtered system~\eqref{eq:general-system-2} that is not an equilibrium of the 
unfiltered system~\eqref{eq:unfiltered-system}
as an
\textit{undesirable equilibrium}.

Our goal is to provide  answers to the following questions: 
\begin{enumerate}
    \item How  do the equilibria of~\eqref{eq:general-system-2} depend on the choice of the pair $(h,\alpha)$?
    \item 
    How do the dynamical properties of~\eqref{eq:general-system-2}  depend on the pair $(h,\alpha)$?  
\end{enumerate}
Before proceeding, we note that under Assumptions~\ref{as:diff-convexity} and~\ref{as:constraint-qualif},~\cite[Theorem 5.3]{AVF-JK:85}
shows that $v_{(h,\alpha)}(\bx)$ and $v(\bx)$ are continuous for all $\bx\in\Sc$. Therefore,~\eqref{eq:unfiltered-system} and \eqref{eq:general-system-2} have a solution for every initial condition in $\Sc$. However, since $v_{(h,\alpha)}$ and $v$ are in general only continuous, but not locally Lipschitz, this solution might not be unique.
However, under different concrete instantiations of~\eqref{eq:unfiltered-system} and~\eqref{eq:general-system-2} considered throughout the paper,
Assumptions~\ref{as:diff-convexity} and~\ref{as:constraint-qualif} are sufficient to guarantee that the corresponding $v_{(h,\alpha)}$ and $v$ are locally Lipschitz.
We refer the interested reader to~\cite{PM-AA-JC:24-ejc} for a survey on regularity properties of such optimization-based controllers.

\section{Impact of CBF on Dynamics and Equilibria}\label{sec:existence-eq-choice-CBF}

In this section, we start examining how the choice of the pair $(h,\alpha)$ affects the dynamics of the filtered system~\eqref{eq:general-system-2}. In particular, we show that:  (i) the dynamics of the filtered system are equal to the dynamics of the unfiltered system around the points where the CBF condition is satisfied strictly (Proposition~\ref{prop:indep-cbf-points-where-cbf-constraint-inactive});
(ii) given a pair $(h,\alpha)$, a point $\bx_*\in\text{Int}(\Sc)$ is an equilibrium of the unfiltered system if and only if it is an equilibrium of the filtered system (Proposition~\ref{prop: effect of CBF on in-eq}); (iii) 
if a point $\bx_*\in\partial\Sc$ is an equilibrium of the unfiltered system, it is also an equilibrium of the filtered system (Proposition~\ref{prop: effect of CBF on boundary-eq}); and (iv) given two pairs $(h_1,\alpha_1)$ and $(h_2,\alpha_2)$, one has that
$v_{(h_1,\alpha_1)}(\bx_*) = v_{(h_2,\alpha_2)}(\bx_*)$ for all $\bx_*\in\partial\Sc$ (Proposition~\ref{prop:boundary-indep-h-alpha-pair}). These results showcase to what extent the properties of equilibria of~\eqref{eq:general-system-2} are independent of the choice of the pair $(h,\alpha)$.

We note that the results presented in this section do not require $v(\bx)$ or $v_{(h,\alpha)}(\bx)$ to be locally Lipschitz, or the solutions of the filtered or unfiltered systems to be unique. 

\begin{proposition}
\longthmtitle{Dynamics are Independent of CBF at Points where CBF Constraint is not Active}\label{prop:indep-cbf-points-where-cbf-constraint-inactive}
   Let
    $(h,\alpha)$ be a strict CBF pair of $\Sc$   
     such that Assumptions~\ref{as:diff-convexity} and~\ref{as:constraint-qualif} hold.
    Let $\bx_*\in\Sc$ and assume that one of the following holds:
    \begin{enumerate}[(i)]
        \item\label{it:first} $\nabla h(\bx_*)^\top ({f}(\bx_*)+g(\bx_*)v_{(h,\alpha)}(\bx_*) )+\alpha(h(\bx_*))>0$;
        \item\label{it:second} $\nabla h(\bx_*)^\top ({f}(\bx_*)+g(\bx_*)v(\bx_*) )+\alpha(h(\bx_*))>0$.
    \end{enumerate}
    Then, there exists an open neighborhood $N_{\bx_*}$ of $\bx_*$ such that $v_{(h,\alpha)}(\bx) = v(\bx)$ for all $\bx\in N_{\bx_*}$.
\end{proposition}
\vspace{-0.5cm}
\begin{pf}
We only prove the case when (\ref{it:first}) holds and note that the argument for (\ref{it:second}) proceeds similarly.
Since $(h,\alpha)$ satisfies Assumptions~\ref{as:diff-convexity} and~\ref{as:constraint-qualif}~\cite[Theorem~5.3]{AVF-JK:85} ensures that $v_{(h,\alpha)}$ is continuous at $\bx_*$.
Moreover, since $\nabla h$, ${f}$ and $g$ are continuous, there exists an open neighborhood $N_{\bx_*}$ of $\bx_*$ such that 
\begin{align}\label{eq:aux}
    &\nabla h(\bx)^\top ({f}(\bx)+g(\bx)v_{(h,\alpha)}(\bx))+\alpha(h(\bx))>0
\end{align}
for all $\bx\in N_{\bx_*}$.
Now, since again $(h,\alpha)$ satisfies Assumptions~\ref{as:diff-convexity} and~\ref{as:constraint-qualif}, the optimizers of~\eqref{eq:general-v-problem} satisfy the Karush–Kuhn–Tucker (KKT) conditions for all $\bx\in N_{\bx_*}$. Let $\lambda:N_{\bx_*}\to\real^q $ be the Lagrange multiplier  associated with the constraints $\ell_i(\bx,\bu,\boldsymbol{\delta})\leq 0$,
and let $\lambda_{q +1}:N_{\bx_*}\to\real$ be the Lagrange multiplier associated with the constraint $F_{h,{f}}(\bx)+\nabla h(\bx)^\top g(\bx)\bu \geq 0$.
Therefore, for all $\bx\in N_{\bx_*}$, the tuple $(v_{h,\alpha}(\bx),\boldsymbol{\delta}_{(h,\alpha)}, \lambda(\bx), \lambda_{ q + 1}(\bx) )$
is a KKT point of~\eqref{eq:general-v-problem}.
Now, from~\eqref{eq:aux}, it follows that $\lambda_{q+1}(\bx)=0$ for all $\bx\in N_{\bx_*}$. 
This implies that the tuple
$(v_{h,\alpha}(\bx),\boldsymbol{\delta}_{(h,\alpha)}(\bx), \lambda(\bx) )$
is a KKT point of problem~\eqref{eq:v-w/o-cbf} for all $\bx\in N_{\bx_*}$,
and therefore $v(\bx) = v_{(h,\alpha)}(\bx)$ for all $\bx\in N_{\bx_*}$.
\hfill $\Box$
\end{pf}

Proposition~\ref{prop:indep-cbf-points-where-cbf-constraint-inactive} shows that at points where the CBF constraint is not active, the filtered and unfiltered systems are equivalent.
Next, we examine the effect of the CBF constraint on equilibria, starting with those in the interior on the safe set.

\begin{proposition}\longthmtitle{Effect of CBF constraint on Interior Equilibria}\label{prop: effect of CBF on in-eq}
   Let
    $(h,\alpha)$ be a strict CBF pair of $\Sc$   
     such that Assumptions~\ref{as:diff-convexity} and~\ref{as:constraint-qualif} hold.
    Then,  a point $\bx_*\in\text{Int}(\Sc)$ is an equilibrium of  the unfiltered system~\eqref{eq:unfiltered-system} if and only if $\bx_*$ is an equilibrium of the filtered system~\eqref{eq:general-system-2}, 
    in which case there exists a neighborhood $N_{\bx_*}$ of $\bx_*$ such that 
    $v_{(h,\alpha)}(\bx)=v(\bx)$, for all
    $\bx\in N_{\bx_*}$.
\end{proposition}
\vspace{-0.5cm}
\begin{pf}
If $\bx_*$ is an equilibrium point of the filtered system~\eqref{eq:general-system-2}, then ${f}(\bx_*)+g(\bx_*)v_{(h,\alpha)}(\bx_*) = 0$. Therefore, from $\bx_*\in\text{Int}(\Sc)$, it follows that $\nabla h(\bx_*)^\top ({f}(\bx_*)+g(\bx_*)v_{(h,\alpha)}(\bx_*) ) + \alpha(h(\bx_*)) > 0$. From Proposition~\ref{prop:indep-cbf-points-where-cbf-constraint-inactive}, it follows that there exists a neighborhood $N_{\bx_*}$ of $\bx_*$ such that $v_{(h,\alpha)}(\bx) = v(\bx)$ for all $\bx\in N_{\bx_*}$. This implies that $\bx_*$ is an equilibrium of the unfiltered system. The reverse implication follows a similar argument.
\hfill $\Box$
\end{pf}

Proposition~\ref{prop: effect of CBF on in-eq} shows that in the interior of the safe set, the equilibria of the filtered and unfiltered systems coincide, for any pair $(h,\alpha)$. In addition, the vector field in a neighborhood of these interior equilibria is the same.  Consequently, undesirable equilibria can only appear on the boundary of the safe set, to which we turn our attention next.

\begin{proposition}\longthmtitle{Effect of CBF Constraint on Boundary Equilibria}\label{prop: effect of CBF on boundary-eq}
   Let
    $(h,\alpha)$ be a strict CBF pair of $\Sc$   
     such that Assumptions~\ref{as:diff-convexity} and~\ref{as:constraint-qualif} hold. Then, if a point $\bx_*\in\partial\Sc$ is an equilibrium of the unfiltered system~\eqref{eq:unfiltered-system},
    it is also an equilibrium of the filtered system~\eqref{eq:general-system-2}.
\end{proposition}

\vspace{-0.5cm}
\begin{pf}
Let $(v(\bx_*), \delta(\bx_*), \lambda(\bx_*) )$ be a KKT point of~\eqref{eq:v-w/o-cbf} at $\bx_*$. Since $\bx_*$ is an equilibrium point of \eqref{eq:unfiltered-system}, $f(\bx_*)+g(\bx_*)v(\bx_*)=\textbf{0}_n$. Let us now show that $(v(\bx_*), \delta(\bx_*), \lambda(\bx_*), 0)$ is a KKT point of~\eqref{eq:general-v-problem}, from which it follows that $\bx_*$ is also an equilibrium point of~\eqref{eq:general-v-problem}.
The non-negativity of the Lagrange multipliers of $(v(\bx_*), \delta(\bx_*), \lambda(\bx_*), 0)$ and the satisfaction of complementary slackness follows from the fact that $(v(\bx_*), \delta(\bx_*), \lambda(\bx_*)$ is a KKT point of~\eqref{eq:v-w/o-cbf}. Finally, the primal feasibility of $(v(\bx_*), \delta(\bx_*))$ follows from the fact that $(v(\bx_*), \delta(\bx_*))$ satisfies the constraints in~\eqref{eq:v-w/o-cbf} and since $\bx_*$ is an equilibrium point, $\nabla h(\bx_*)^\top (f(\bx_*) + g(\bx_*)v(\bx_*)) + \alpha( h(\bx_*) ) = \alpha( h(\bx_*) ) = 0$.
\hfill $\Box$ 
\end{pf}
Proposition~\ref{prop: effect of CBF on boundary-eq}
shows that the addition of the CBF constraint in~\eqref{eq:general-v-problem} does not affect the equilibria of the unfiltered system in the boundary of the safe set, again for any pair $(h,\alpha)$. However, the vector field in a neighborhood of these boundary equilibria might change due to the additional  CBF constraint. Additionally, new equilibria might emerge in $\partial\Sc$, which will be undesirable by definition.

The following result establishes that, under some mild assumptions, the vector field in~\eqref{eq:general-system-2} evaluated at points in $\partial\Sc$ is independent of the choice of the pair $(h,\alpha)$.

\begin{proposition}\longthmtitle{Dynamics in the Boundary are Independent of $(h,\alpha)$ Pair}\label{prop:boundary-indep-h-alpha-pair}
   For $i \in \{1,2\}$, let
    $(h_i,\alpha_i)$ be a strict CBF pair of $\Sc$   
     such that Assumptions~\ref{as:diff-convexity} and~\ref{as:constraint-qualif} hold. Then,  $v_{(h_1,\alpha_1)}(\bx) = v_{(h_2,\alpha_2)}(\bx)$ for all $\bx\in\partial\Sc$.
\end{proposition}
\vspace{-0.5cm}
\begin{pf}
Since Assumption~\ref{as:diff-convexity} holds and 
the pairs $(h_1,\alpha_1)$ and $(h_2,\alpha_2)$ satisfy Assumptions~\ref{as:diff-convexity} and~\ref{as:constraint-qualif}, $v_{(h_1,\alpha_1)}(\bx)$ and $v_{(h_2,\alpha_2)}(\bx)$
satisfy the KKT equations of~\eqref{eq:general-v-problem} for $(h_1,\alpha_1)$ and $(h_2,\alpha_2)$, respectively. \\
Let $(v_{(h_1,\alpha_1)}(\bx),\bde_{(h_1,\alpha)}(\bx), \lambda(\bx),\lambda_{q +1}(\bx))$ be a KKT point for~\eqref{eq:general-v-problem} under $(h_1,\alpha_1)$.
If $\bx\in\partial\Sc$, we must have $h_1(\bx) = h_2(\bx) = 0$ and $\nabla h_1(\bx) = \zeta_{(h_1,h_2)}(\bx) \nabla h_2(\bx)$, with $\zeta_{(h_1,h_2)}(\bx) \in \real_{>0}$, from Lemma~\ref{lem: the existence of positive scalar function}. 
This implies that 
\begin{align*}
    (v_{(h_1,\alpha_1)}(\bx),\bde_{(h_1,\alpha)}(\bx), \lambda(\bx),\lambda_{q +1}(\bx)/\zeta_{(h_1,h_2)}(\bx) )
\end{align*}
is a KKT point of~\eqref{eq:general-v-problem} under $(h_2,\alpha_2)$, and therefore $v_{(h_1,\alpha_1)}(\bx) = v_{(h_2,\alpha_2)}(\bx)$.
\hfill $\Box$
\end{pf}

\vspace{-0.4cm}
Since the vector field in~\eqref{eq:general-system-2} evaluated at points in $\partial\Sc$ is independent of the choice of pair $(h,\alpha)$, trajectories that stay in the boundary at all times are not affected by the choice of $(h,\alpha)$. Formally, we have the following result.

\begin{corollary}\longthmtitle{Trajectories along the Boundary are Independent of CBF}\label{cor:trajs-boundary-independent}
 For $i \in \{1,2\}$, let
    $(h_i,\alpha_i)$ be a strict CBF pair of $\Sc$   
     such that Assumptions~\ref{as:diff-convexity} and~\ref{as:constraint-qualif} hold. 
     Let $\tau \in (0,+\infty]$.
    Then, $\by: [0,{\tau} ) \to \partial\Sc$ is a solution of the 
    filtered system under the pair $(h_1,\alpha_1)$ if and only if it 
    is a solution of the filtered system under the pair $(h_2,\alpha_2)$.
    In particular, given $\bx_*\in\partial\Sc$, $\bx_*$ is an equilibrium point of the filtered system under the pair $(h_1,\alpha_1)$ if and only if it is an equilibrium of the filtered system under the pair $(h_2,\alpha_2)$.
\end{corollary}
\vspace{-0.4cm}
\begin{pf}
Since $\by(t)\in\partial\Sc$ for all $t\in[0, {\tau} )$, by Proposition~\ref{prop:boundary-indep-h-alpha-pair}
it follows that $v_{(h_1,\alpha_1)}(\by(t)) = v_{(h_2,\alpha_2)}(\by(t))$ for all $t\in[0, { \tau} )$.
This implies that for all $t\in[0,\tau)$,
\begin{align*}
({f} + gv_{(h_1,\alpha_1)})(\by(t)) = 
({f} + gv_{(h_2,\alpha_2)})(\by(t)),
\end{align*}
and therefore $\by$ is a solution of the filtered system under the pair $(h_1,\alpha_1)$ if and only if it is a solution of the filtered system under the pair $(h_2,\alpha_2)$. Finally, the last claim follows from the fact that if $\bx_*\in\partial\Sc$ is an equilibrium point for the filtered system under pair $(h_1,\alpha_1)$ (resp. $(h_2,\alpha_2)$), then $\by(t) = \bx_*$ for all $t\geq0$ is a valid solution of the filtered system under the pair $(h_1,\alpha_1)$ (resp. $(h_2,\alpha_2)$). 
\hfill $\Box$
\end{pf}
\vspace{-0.5cm}
Corollary~\ref{cor:trajs-boundary-independent} implies that the number and location of limit cycles and equilibria
(and, in particular, undesirable equilibria) 
in $\partial\Sc$ are independent of the pair $(h,\alpha)$. 
We note also that Corollary~\ref{cor:trajs-boundary-independent} does not require uniqueness of trajectories of the closed-loop system.

\begin{remark}\longthmtitle{Extension to Multiple  Obstacles}\label{remark: Extension to multiple  obstacles}
{\rm
    The results in this section can be extended to the case of multiple disjoint obstacles. In particular, assume that the unsafe set can be represented as $\mathcal{S}^c=\bigcup_{i=1}^r \mathcal{S}_i^c$ and $\overline{\mathcal{S}_i^c}\cap\overline{\mathcal{S}_j^c}=\varnothing$
    for all $i\neq j$. Let $(h^{(i)},\alpha^{(i)})$ be a strict CBF pair satisfying Definition~\ref{def:cbf} for the set $\Sc_i$. Then, a generalization of \eqref{eq:general-v-problem} for the case with multiple CBF constraints reads as
\begin{align} \label{eq:general-v-problem-multi-cbf} 
\begin{bmatrix}
    v_{[r]}(\bx) \\ \delta_{[r]}(\bx)
\end{bmatrix}&:=\arg\underset{ \boldsymbol{[\bu; \boldsymbol{\delta}]} \in \mathbb{R}^{m+m_0}}{\min} \ell(\bx,\bu,\boldsymbol{\delta})  \\
\notag
& \text { s.t. }  \ell_i(\bx,\bu,\boldsymbol{\delta} )\leq 0,~\forall~ i = 1,\ldots q \\
\notag
&\qquad F_{h^{(j)},\alpha^{(j)},{f}}(\bx)+\nabla h^{(j)}(\bx)^\top g(\bx)\bu  \geq 0,\\
\notag
&\qquad~\forall~j\in [r].
\end{align}
where $ F_{h^{(j)},\alpha^{(j)},{f}}(\bx) = \nabla h^{(j)}(\bx)^\top {f}(\bx) + \alpha^{(j)}(h^{(j)}(\bx))$. 
We assume that Slater's condition holds for~\eqref{eq:general-v-problem-multi-cbf} for all $\bx\in\mathcal{S}$ and that  $\nabla h^{(j)}(\bx)$ and $\alpha^{(j)}(\cdot)$ are continuously differentiable for all $j\in[r]$.

For equilibria on $\operatorname{int}(\mathcal{S})$, the general strategy consists in viewing the first $r-1$ CBF constraints as  general constraints $\ell_i$ in \eqref{eq:general-v-problem}, applying Proposition~\ref{prop:indep-cbf-points-where-cbf-constraint-inactive}, and repeating the argument $r-1$ times to get a result similar to Proposition~\ref{prop: effect of CBF on in-eq}. For equilibria on $\partial\mathcal{S}$, given any  $\bx_*\in\partial\mathcal{S}$, one must have that there exists a unique $ j\in[r]$ such that $\bx_*\in\partial\mathcal{S}_j$ (since we assume that the obstacles are disjoint). One can then view the $i$-th (for $i = j$) CBF constraint as the CBF constraint in \eqref{eq:general-v-problem} and view the other (for $i \neq j$) CBF constraints
as general constraints~$\ell_i$.  By applying Proposition~\ref{prop:indep-cbf-points-where-cbf-constraint-inactive} to the $(h^{(i)},\alpha^{(i)})$ pairs for $i\neq j$, we get that in a neighborhood of $\bx_*$ the dynamics are independent of the pairs $(h^{(i)},\alpha^{(i)})$ for $i\neq j$. Moreover, by Corollary~\ref{cor:trajs-boundary-independent} the existence of $\bx_*$ is independent of the pair $(h^{(j)},\alpha^{(j)})$.
We omit the details due to space limitations. \hfill $\Box$
}
\end{remark}

\vspace{-0.3cm}

\section{Case of Safety Filters: Impact of CBF Selection}\label{sec:choice-safety-filters}
\vspace{-0.3cm}
In this section, we study the particular case of~\eqref{eq:general-v-problem} that arises when filtering a nominal controller $k:\real^n \rightarrow \real^m$ with a CBF constraint. Given a nominal controller $k:\real^n\to\real^m$ (which often is assumed to  stabilize a certain desirable equilibrium $\bx^*$), safety filters seek to find the control input that is closest to $k$ (with a distance induced by a positive definite matrix-valued function $G:\real^n\to\real^{m\times m}$) and satisfies the CBF constraint.
This leads to the closed-loop system $\dot \bx={f}(\bx)+g(\bx) k(\bx) + g(\bx)\breve{v}_{(h,\alpha)}(\bx)  $ with $\breve{v}_{(h,\alpha)}$ defined as follows:
\begin{align} \label{eq:v-safety-filters} 
    &\breve{v}_{(h,\alpha)}(\bx)=\arg\underset{ \bu \in \mathbb{R}^n}{\min}  \| \bu \|_{G(\bx)}^2 \\
    \notag
    &\text { s.t. } \nabla h(\bx)^\top ({f}(\bx)+g(\bx) (k(\bx) + \bu ) ) + \alpha(h(\bx)) \geq 0.
\end{align}
We note that~\eqref{eq:v-safety-filters} is a particular case of~\eqref{eq:general-v-problem}, obtained by letting $m_0=0$, $q=0$ and $l(\bx,\bu,\boldsymbol{\delta})=\frac{1}{2}\norm{\bu}_{G(\bx)}^2$. Therefore, all the results obtained in Section~\ref{sec:existence-eq-choice-CBF} apply to this case. Here, we are interested in going further to draw conclusions on
whether the CBF pair affects the stability properties of the undesirable equilibria of~\eqref{eq:general-system-2}. To tackle this, we study how the Jacobian matrix depends on the choice of the pair $(h,\alpha)$.

By~\cite[Theorem 1]{WSC-DVD:20}, 
 $\breve{v}_{(h,\alpha)}$ can be written explicitly as

\begin{align}\label{eq:v-safety-filter-explicit}
    \breve{v}_{(h,\alpha)}(\bx) = \begin{cases}
        \textbf{0}_m, &\ \text{if} \ \eta(\bx) \geq 0, \\
        \bar{\bu}(\bx), &\ \text{if} \ \eta(\bx) < 0,
    \end{cases}
\end{align}
where $\bar{\bu}(\bx):=-\frac{\eta(\bx) G(\bx)^{-1}g(\bx)^\top\nabla h(\bx) }{\| 
 g(\bx)^\top \nabla h(\bx)  \|_{G^{-1}(\bx)}^2}$
 and $\eta(\bx):=\nabla h(\bx)^\top ( {f}(\bx) + g(\bx)k(\bx) ) + \alpha(h(\bx))$. If $(h,\alpha)$ is a strict CBF pair,  problem~\eqref{eq:v-safety-filters} satisfies Slater's condition for all $\bx\in\Sc$. 
 . It follows, using
 \cite[Proposition 3.3]{PM-AA-JC:24-ejc},  that if $f$, $g$, $k$, $G$, $\nabla h$ and $\alpha$ are locally Lipschitz, then $\breve{v}_{(h,\alpha)}$ is also locally Lipschitz.

We note that~\cite[Lemma 1]{YC-PM-EDA-JC:24-cdc} provides a characterization of the equilibria set $\Ec_{\text{sf}}$ of the closed-loop system $\dot{\bx} = \tilde{f}(\bx) + g(\bx)\breve{v}_{(h,\alpha)}(\bx)$ (where $\tilde{f}(\bx) := f(\bx) + g(\bx)k(\bx)$) as $\Ec_{\text{sf}}=\hat{\Ec}_{\text{sf}}\cup\{\bx:~\tilde{f}(\bx)=0\}$,
\begin{equation}\label{eq: safety filter_underisable eq}
    \hat{\Ec}_{\text{sf}} := \{\bx\in\partial\mathcal{S}:~\exists~\delta<0  \text{ s.t. } {\tilde{f}(\bx)}=\delta D(\bx)\nabla h(\bx)\},
\end{equation}
and $D(\bx) := g(\bx) G^{-1}(\bx) { g(\bx)^{\top}}$.
Note that under the assumptions of Corollary~\ref{cor:trajs-boundary-independent}, the set $\hat{\Ec}_{\text{sf}}$ is independent of the pair~$(h,\alpha)$. 

Next, let us study the stability properties of the equilibria of the closed-loop system $\dot{\bx} = \tilde{f}(\bx) + g(\bx)\breve{v}_{(h,\alpha)}(\bx)$.
First, let $\bx^*$ be a desirable equilibrium of system $\dot \bx=\tilde{f}(\bx)$.  By Proposition~\ref{prop: effect of CBF on in-eq}, if $\bx^* \in \{ \bx:~{\tilde{f}(\bx)}=0\}\subset\text{Int}(\Sc)$, the closed-loop system in a neighborhood of $\bx^*$ is independent of the $(h,\alpha)$ pair. Hence, the stability properties of $\bx^*$ are also independent of the $(h,\alpha)$ pair. 

Now, let us turn our attention to the equilibria in $\hat{\Ec}_{\text{sf}}$.
To determine the stability of the undesirable equilibria, we investigate the Jacobian of $\tilde{f}(\bx) + g(\bx)\breve{v}_{(h,\alpha)}(\bx)$.

\begin{proposition}\longthmtitle{Jacobian Evaluation at Undesirable Equilibria of Safety Filters}\label{prop:jacobian-evaluation-choice-cbf-safety-filters}
Let $(h,\alpha)$ be a strict CBF pair.
Suppose that $\tilde{f}$, $g$, $G$ and $\alpha$ are continuously differentiable, and $h$ is twice continuously differentiable. For any undesirable equilibrium $\bx_*\in\hat{\Ec}_{\text{sf}}$, the Jacobian of $\tilde{f}(\bx) + g(\bx)\breve{v}_{(h,\alpha)}(\bx)$ evaluated at $\bx_*$ is 
\begin{align}
\label{eq: Jacobian for safety filter}
   J_{h,\alpha} \mid_{\bx_*}=&\Big[ J_{{\tilde{f}}}-\frac{ D \nabla h \nabla h^\top}{\nabla h^\top D \nabla h  } [ J_{{\tilde{f}}}+\alpha^\prime(0)\textbf{I}_n ] \\
\notag
&- \frac{D[ \nabla h^\top {\tilde{f}} \textbf{I}_n -  \nabla h {\tilde{f}}^\top ]H_{h}}{\nabla h^\top D \nabla h } 
 \\
 \notag
 &- \frac{ \nabla h^\top \tilde{f}[\nabla h^\top D \nabla h \textbf{I}_n - D \nabla h \nabla h^\top] \frac{\partial D}{\partial\bx}\nabla h}{(\nabla h^\top D \nabla h)^2 }   \Big] \bigg\rvert_{\bx_*},
\end{align}
where we have dropped the dependency in $\bx$ for simplicity, $J_{{\tilde{f}}}(\bx_*)$ is the Jacobian matrix of ${\tilde{f}(\bx)}$ evaluated at $\bx_*$, and $H_{h}(\bx_*)$ is the Hessian of $h(\bx)$ evaluated at $\bx_*$. In addition, it holds that
\begin{equation}\label{eq: gradient is an eigenvector}   
    \nabla h(\bx_*)^\top J_{h,\alpha}\mid_{\bx_*}=-\alpha^\prime(0) \nabla h(\bx_*)^\top.
\end{equation}
\end{proposition}
\begin{pf}
For any undesirable equilibrium $\bx_*$, we have $\eta(\bx_*)<0$. By continuity of $\eta(\bx)$, there exists a neighborhood $U(\bx_*)$ such that $\eta(\bx)<0$ for all $\bx\in U(\bx_*)$.
By \eqref{eq:v-safety-filter-explicit}, one can write $\tilde{f}(\bx) + g(\bx)\breve{v}_{(h,\alpha)}(\bx)=\tilde{f}(\bx)-\frac{\eta(\bx) D(\bx)\nabla h(\bx) }{\nabla h(\bx)^\top D(\bx )\nabla h(\bx)  }$ for any $\bx \in U(\bx_*)$, where we recall that $D(\bx) = g(\bx) G^{-1}(\bx) { g(\bx)^{\top}}$. Then the Jacobian of $\tilde{f}(\bx)+g(\bx)\breve{v}{(h,\alpha)}(\bx)$ at $\bx_*$ is given by
$J_{h,\alpha} \mid_{\bx_*}=\frac{\partial}{\partial\bx} \Big[\tilde{f}(\bx)-\frac{\eta(\bx) D(\bx)\nabla h(\bx) }{\nabla h(\bx)^\top D(\bx )\nabla h(\bx)  }\Big]\bigg\rvert_{\bx_*}$,  resulting in \eqref{eq: Jacobian for safety filter}. Next, by~\eqref{eq: safety filter_underisable eq}, it holds that $\nabla h(\bx_*)^\top D(\bx_*) [\nabla h(\bx_*)^\top {\tilde{f}}(\bx_*)] = {\tilde{f}}(\bx_*)^\top[\nabla h(\bx_*)^\top D(\bx_*) \nabla h(\bx_*) ]$, which together with \eqref{eq: Jacobian for safety filter} implies \eqref{eq: gradient is an eigenvector}. 
\hfill $\Box$
\end{pf}
\vspace{-0.3cm}
By \eqref{eq: gradient is an eigenvector}, it follows that $-\alpha_1^\prime(0)$ is always an eigenvalue of the Jacobian evaluated at any undesirable equilibrium. The following result shows that if $h_1 \stackrel{\text{H}}{\sim} h_2$, the rest of $n-1$ eigenvalues are independent of the choice of CBF pair $(h,\alpha)$.

\begin{proposition}\longthmtitle{Jacobian at Undesirable Equilibria of Safety Filters as a Function of $(h,\alpha)$}\label{prop:jacobian-choice-cbf-safety-filters}
Let $(h_1,\alpha_1)$, $(h_2,\alpha_2)$ be two strict CBF pairs.
Suppose that $\tilde{f}$, $g$, $G$, $\alpha_1$ and $\alpha_2$ are continuously differentiable, and $h_1$, and $h_2$ are twice continuously differentiable. 
If $h_1 \stackrel{\text{H}}{\sim} h_2$,  then for any $\bx_*\in \hat{\Ec}_{\text{sf}}$,
\begin{align*}
&\frac{\operatorname{det}\left(sI-J_{h_2,\alpha_2}\mid_{\bx_*}\right)}{s+\alpha_2^\prime(0)}=\frac{ \operatorname{det}\left(sI-J_{h_1,\alpha_1}\mid_{\bx_*}\right)}{s+\alpha_1^\prime(0)}.
\end{align*}
\end{proposition}

\vspace{-0.7cm}
\begin{pf}
Since $\nabla h_1^\top D \nabla h_1>0$, 
for $\bx_*\in\hat{\Ec}_{\text{sf}}$, then $D(\bx_*) \nabla h_1(\bx_*)\neq \textbf{0}_n^\top$, and therefore there exist $\xi_i$, for $i=2,...,n$ such that $\left\{\frac{D(\bx_*) \nabla h_1(\bx_*)}{\|D(\bx_*) \nabla h_1(\bx_*)\|} \right\}\cup\{\xi_i\}_{i=2}^n$
is an orthonormal basis of $\mathbb{R}^n$ (since $\nabla h_1(\bx_*)^\top D(\bx_*) \nabla h_1(\bx_*)\neq0$, it also follows that $\{\nabla h_1(\bx_*)\}\cup\{\xi_i\}_{i=2}^n$ is a basis of $\mathbb{R}^n$).
Using the orthogonality of the basis, we have that for $i=2,...,n$,
\begin{equation}\label{eq:compuation of charateristic polynomial}
\left(J_{h_1,\alpha_1}\mid_{\bx_*}\right)^\top \xi_i=C_{h_1,\bx_*}^\top \xi_i
\end{equation}
where $C_{h_1,\bx_*}:=J_{\tilde{f}} - \frac{\nabla h_1^\top {\tilde{f}}D  H_{h_1}}{\nabla h_1^\top D \nabla h_1 }  - \frac{ \nabla h_1^\top \tilde{f}\nabla h_1^\top D \nabla h_1 \frac{\partial D}{\partial\bx}\nabla h_1}{(\nabla h_1^\top D \nabla h_1)^2 }
     $ is evaluated at $\bx_*$.
The right-hand side of \eqref{eq:compuation of charateristic polynomial} can be written as a linear combination of $\{\nabla h_1(\bx_*)\}\cup\{\xi_j\}_{j=2}^n$ as $C_{h_1,\bx_*}^\top \xi_j= M_{1,j} \nabla h_1(\bx_*)+\sum_{j=2}^n M_{i,j} \xi_j$.

\vspace{-0.2cm}
Next, define $T:= \begin{bmatrix}
       \nabla h_1(\bx_*),~ \xi_2,\cdots,\xi_n
    \end{bmatrix}\in\mathbb{R}^{n\times n}$, $\gamma_1 := \begin{bmatrix}
       M_{1,2} &  M_{1,3} & \cdots & M_{1,n}
\end{bmatrix}\in\mathbb{R}^{1\times (n-1)}$, a matrix $M= [M_{i+1,j+1}]\in\real^{(n-1)\times(n-1)}$, where the entry at the $i$-th row and $j$-th column of $M$ is equal to $M_{i+1,j+1}$.  Then we can write
$
  T^{-1}  \left(J_{h_1,\alpha_1}\mid_{\bx_*}\right)^\top T=\begin{bmatrix}
      -\alpha_1^\prime(0) & \gamma_1\\
      \zero_{n-1} & M
  \end{bmatrix}.
$

Next, we compute $ T^{-1}  \left(J_{h_2,\alpha_2}\mid_{\bx_*}\right)^\top T$. First, note that 
\begin{align*}
    &\nabla h_1(\bx_*)^\top J_{h_2,\alpha_2}\mid_{\bx_*}=\frac{1}{\zeta(\bx_*)}\nabla h_2(\bx_*)^\top J_{h_2,\alpha_2}\mid_{\bx_*} \\
    &=-\frac{1}{\zeta(\bx_*)}\alpha_2^\prime(0) \nabla h_2(\bx_*)^\top=-\alpha_2^\prime(0) \nabla h_1(\bx_*)^\top,
\end{align*}
where we use the fact that $\nabla h_2(\bx_*)=\zeta(\bx_*)\nabla h_1(\bx_*)$.

 Next, we compute $\xi_j^\top J_{h_2,\alpha_2}\mid_{\bx_*}$. Since we have already computed the value of $\left(J_{h_1,\alpha_1}\mid_{\bx_*}\right)^\top \xi_j$, cf.~\eqref{eq:compuation of charateristic polynomial}, it suffices to compute $\left( J_{h_2,\alpha_2}\mid_{\bx_*}-J_{h_1,\alpha_1}\mid_{\bx_*}\right)^\top \xi_j$.

 Consider two equivalent CBFs such that $\nabla h_2(\bx_*)=\zeta(\bx_*)\nabla h_1(\bx_*)$, $H_{h_2}(\bx_*) = \nabla h_1(\bx_*) \tilde{\zeta}(\bx_*)^\top+ \tilde{\zeta}(\bx_*) \nabla h_1(\bx_*)^\top $ $ +\zeta(\bx_*) H_{h_1}(\bx_*)$, we use the expression for the Jacobian in~\eqref{eq: Jacobian for safety filter} to get for any $\bx_*\in\hat{\Ec}_{\text{sf}}$ that
 \vspace{-0.3cm}
\begin{align}
\notag
&J_{h_2,\alpha_2}\mid_{\bx_*}- J_{h_1,\alpha_1}\mid_{\bx_*} =\Big[-( \alpha_2^\prime(0)-\alpha_1^\prime(0) )\frac{ D \nabla h_1 \nabla h_1^\top}{\nabla h_1^\top D \nabla h_1  }\\
\notag
&\qquad\qquad+\frac{D  \nabla h_1 {\tilde{f}}^\top H_{h_1}}{\nabla h_1^\top D \nabla h_1 }-\frac{D  \nabla h_1 {\tilde{f}}^\top H_{h_2}}{\nabla h_2^\top D \nabla h_1 }  \\
&\qquad\qquad+\frac{\nabla h_2^\top {\tilde{f}} D  H_{h_2}}{\nabla h_2^\top D \nabla h_2 }-\frac{\nabla h_1^\top {\tilde{f}} D  H_{h_1}}{\nabla h_1^\top D \nabla h_1 }\Big]\bigg\rvert_{\bx_{*}} . \label{eq: difference between Jaconbian-step1}
\end{align}
We note that the last two terms in \eqref{eq: difference between Jaconbian-step1} yield
 \vspace{-0.3cm}
\begin{equation}\label{eq: difference between Jaconbian-step2}
\begin{aligned}
    &\frac{\nabla h_2^\top {\tilde{f}} D  H_{h_2}}{\nabla h_2^\top D \nabla h_2 }-\frac{\nabla h_1^\top {\tilde{f}} D  H_{h_1}}{\nabla h_1^\top D \nabla h_1 }\\
=&\frac{ \nabla h_2^\top {\tilde{f}} }{\nabla h_2^\top D \nabla h_2 } D \nabla h_1 \tilde{\zeta}^\top - \frac{ \nabla h_2^\top {\tilde{f}}}{\nabla h_2^\top D \nabla h_2 }D \tilde{\zeta} \nabla h_1^\top .
\end{aligned}
\end{equation}
The combination of~\eqref{eq: difference between Jaconbian-step1} and~\eqref{eq: difference between Jaconbian-step2} yields the following
\vspace{-0.1cm}
\begin{align}
\notag
&J_{h_2,\alpha_2}\mid_{\bx_*}- J_{h_1,\alpha_1}\mid_{\bx_*} =\Big[-( \alpha_2^\prime(0)-\alpha_1^\prime(0) )\frac{ D \nabla h_1 \nabla h_1^\top}{\nabla h_1^\top D \nabla h_1  }\\
\label{eq:Jacobian-equiv}
&~~~+\frac{D  \nabla h_1 {\tilde{f}}^\top H_{h_1}}{\nabla h_1^\top D \nabla h_1 }-\frac{D  \nabla h_1 {\tilde{f}}^\top H_{h_2}}{\nabla h_2^\top D \nabla h_1 }  \\
\notag
&~~~+\frac{ \nabla h_2^\top {\tilde{f}} }{\nabla h_2^\top D \nabla h_2 } D \nabla h_1 \tilde{\zeta}^\top- \frac{ \nabla h_2^\top {\tilde{f}}}{\nabla h_2^\top D \nabla h_2 }D \tilde{\zeta} \nabla h_1^\top\Big]\bigg\rvert_{\bx_{*}} .
\end{align}

\vspace{-0.4cm}
Using~\eqref{eq:Jacobian-equiv}, the computation of $\left( J_{h_2,\alpha_2}\mid_{\bx_*}\!-J_{h_1,\alpha_1}\mid_{\bx_*}\right)^\top \xi_j$ boils down to computing the multiplication between $\xi_j^\top$ and the five terms in \eqref{eq:Jacobian-equiv}. Since $\xi_j^\top D(\bx_*)\nabla h_1(\bx_*)=0$ , it follows that the multiplication between $\xi_j^\top$ and first four terms are $0$. 
Therefore, for any $i=2,...,n$, we have 
\vspace{-0.3cm}
\begin{align*}
    &\left( J_{h_2,\alpha_2}\mid_{\bx_*}-J_{h_1,\alpha_1}\mid_{\bx_*}\right)^\top \xi_j=- \frac{ \nabla h_2^\top {\tilde{f}}}{\nabla h_2^\top D \nabla h_2 }(\tilde{\zeta}^\top D \xi_j)   \nabla h_1.
\end{align*}
Hence $\left(J_{h_2,\alpha_2}\mid_{\bx_*}\right)^\top \xi_j=(M_{1,j}+\beta_{2,j}) \nabla h_1+\sum_{j=2}^n M_{i,j} \xi_j,$
where $\beta_{2,j}:=-\frac{ \nabla h_2^\top {\tilde{f}}}{\nabla h_2^\top D \nabla h_2 }\tilde{\zeta}^\top D \xi_j$. 

Now, define $\gamma_2:=\begin{bmatrix}
    \beta_{2,2} &  \beta_{2,3} & \cdots  &\beta_{2,n}
\end{bmatrix}\in\mathbb{R}^{1 \times (n-1)}$, and note that it holds that
$
  T^{-1}  \left(J_{h_2,\alpha_2}\mid_{\bx_*}\right)^\top T=\begin{bmatrix}
      -\alpha_2^\prime(0) & \gamma_1+\gamma_2\\
      \zero_{n-1} & M
  \end{bmatrix} .
$
Hence, we have that $\frac{\operatorname{det}\left(sI-J_{h_2,\alpha_2}\mid_{\bx_*}\right)}{s+\alpha_2^\prime(0)} $  $   = \operatorname{det}\left(sI-M\right)=  \frac{ \operatorname{det}\left(sI-J_{h_1,\alpha_1}\mid_{\bx_*}\right)}{s+\alpha_1^\prime(0)}.$
\hfill $\Box$ \end{pf}
\vspace{-0.5cm}
Proposition~\ref{prop:jacobian-choice-cbf-safety-filters} shows that the stability properties under the safety filter of the undesirable equilibria are the same for all equivalent CBFs. This aligns with the findings of Section~\ref{sec:existence-eq-choice-CBF}: the equilibria of unfiltered system are preserved (Proposition~\ref{prop: effect of CBF on in-eq} and~\ref{prop: effect of CBF on boundary-eq}); and the  vector field and trajectories at the boundary are independent of the CBF pair (Proposition \ref{prop:boundary-indep-h-alpha-pair} and Corollary~\ref{cor:trajs-boundary-independent}).  Overall, the local closed-loop system behavior around the equilibria is determined by the geometry of the safe set rather than by a specific analytical parameterization of the equivalent CBFs. The same conclusions extend to multiple CBF constraints, as discussed in Remark~\ref{remark: Extension to multiple  obstacles}.

\vspace{-0.2cm}
\section{Numerical Simulation}\label{sec:simulations}

\begin{figure}[t!]
  \centering 
  {\includegraphics[width=0.48\textwidth]{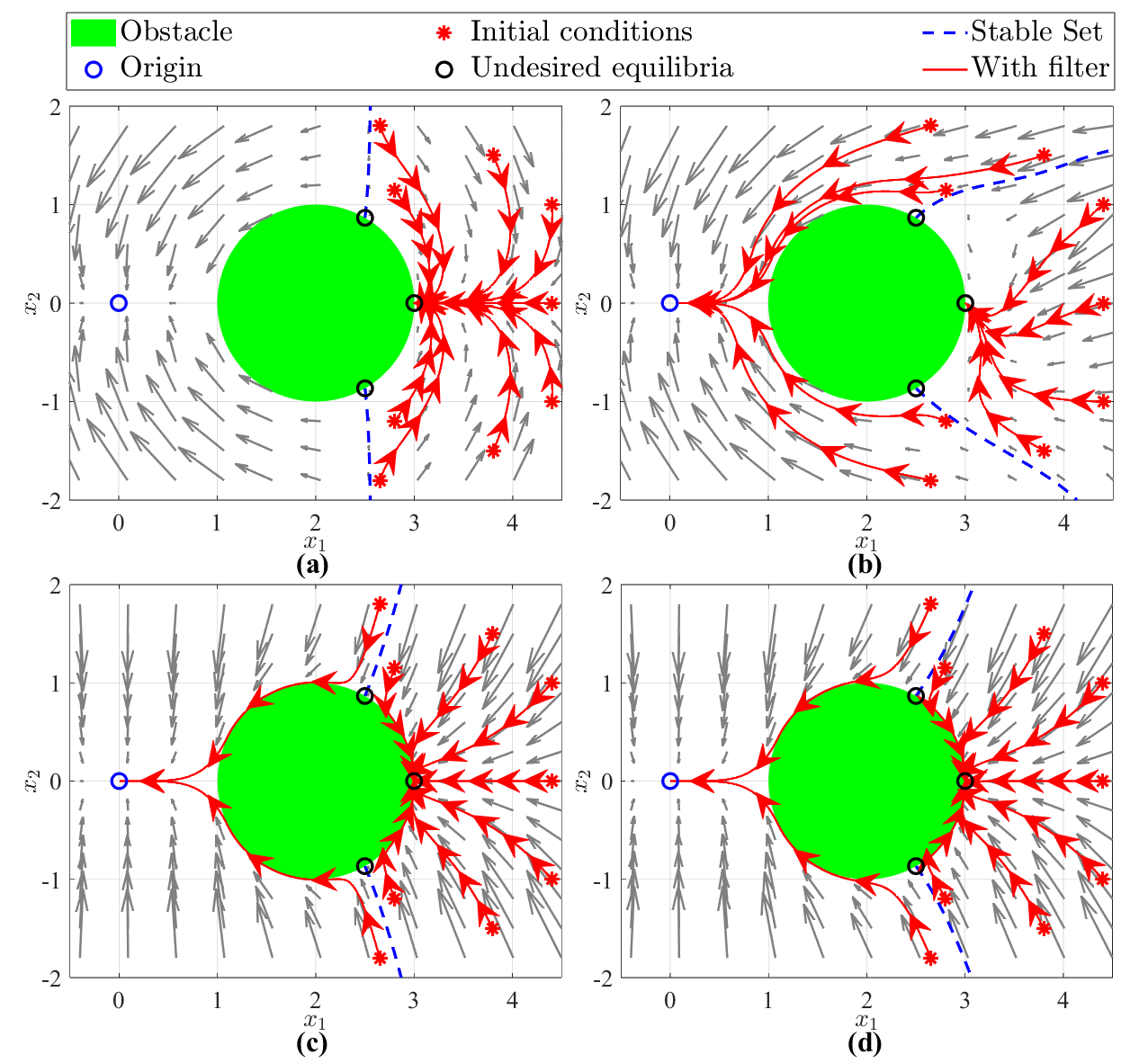}} 
   \vspace{-.4cm}
  \caption{Examples of
trajectories (red arrowed curves) of an LTI planar system with a safety filter for a circular obstacle; the figures also show the vector fields (gray arrows), the undesired equilibria, the stable sets (stable manifolds) of the saddle equilibria, and the desired equilibrium (which is the origin). The plots are generated with different CBF pairs:  (a): with CBF pair $(h_1,\alpha_1)$;  (b): with CBF pair $(h_2,\alpha_1)$; (c): with CBF pair $(h_1,\alpha_2)$; (d) with CBF pair $(h_2,\alpha_2)$.  }
  \label{fig: fully actuated cases}
   \vspace{.1cm}
\end{figure}

This section illustrates our results in a numerical example. We also study qualitatively the dependence of other dynamical properties, like transient behavior or the size of the region of attraction of the desired equilibria, on the CBF pair.
Consider the integrator dynamics $\dot \bx=\bu$ with nominal controller
$k(\bx)=\operatorname{diag}([-1,-5])\bx$,    
which globally asymptotically stabilizes the origin. There is an obstacle in the environment in the form of a ball centered at $(2,0)$ of radius one, so let $\mathcal{S} = \{\bx:~\|\bx-(2,0)^\top\|^2-1\geq 0 \}$ be the safe set.
We represent this with two CBF pairs, given by $h_1(\bx):=\|\bx-(2,0)^\top\|^2-1$, $h_2(\bx):=(\|\bx-(5,1)^\top\|^2+1)h_1(\bx)$, $\alpha_1(s):=s$ and $\alpha_2(s)=10s$. Note that $h_1\stackrel{\text{H}}{\sim}h_2$.
In Figure~\ref{fig: fully actuated cases}, we apply the safety filter with CBF pairs $(h_i,\alpha_j)$, $i,j\in\{1,2\}$ (here, $G(\bx)=g(\bx)^\top g(\bx)=\textbf{I}_2$) to the unfiltered system $\dot \bx=k(\bx)$.

We compute the undesirable equilibria using the pair $(h_1,\alpha_1)$, as outlined in Section~\ref{sec:choice-safety-filters}, and obtain three: $(\frac{5}{2},\frac{\sqrt{3}}{2})^\top$, $(\frac{5}{2},-\frac{\sqrt{3}}{2})^\top$ and $(3,0)^\top$. By Corollary~\ref{cor:trajs-boundary-independent}, these equilibria exist for any CBF pair.
Furthermore, using the expression of the Jacobian in Proposition~\ref{prop:jacobian-choice-cbf-safety-filters} with pair $(h_1,\alpha_1)$, we deduce that the equilibrium $(3,0)^\top$ is asymptotically stable and the other two are saddle points. By Proposition~\ref{prop:jacobian-choice-cbf-safety-filters}, these stability properties hold for any pair for which the CBF is in the same equivalence class as~$h_1$.
This independence can be observed in Figure~\ref{fig: fully actuated cases} (recall that $h_1\stackrel{\text{H}}{\sim}h_2$). In each of the plots,  the gray arrows represent the vector field and the red arrow lines
represent the trajectories.
The trajectories converging to  $(\frac{5}{2},\frac{\sqrt{3}}{2})^\top$, $(\frac{5}{2},-\frac{\sqrt{3}}{2})^\top$ and the points in the boundary of the obstacle with a value of $x_1$ smaller than or equal to $\frac{5}{2}$ constitute the boundary of the region of attraction of the origin.

Even though our results show that the number, location, and local stability properties of the equilibria remain unchanged under different CBF pairs, other dynamical properties of the filtered system may indeed change with the CBF pair.
For example, Figure~\ref{fig: fully actuated cases} shows that the size of the region of attraction of the origin is dependent on the CBF pair.
We analyze the effect of the CBF and the extended class-$\Kc_{\infty}$ function on the region of attraction of the origin next. 
(i) If we fix $h=h_1$ (Figure~\ref{fig: fully actuated cases}(a),(c)), the region of attraction of the origin is larger for $\alpha_2$ (which has a larger slope) than for $\alpha_1$.
However, if we fix $h=h_2$ (Figure~\ref{fig: fully actuated cases}(b),(d)), the region of attraction of the origin is larger for $\alpha_1$ (which has a smaller slope) than for $\alpha_2$.
This suggests a complex dependence of the dynamical properties such as the region of attraction of the desired equilibrium on the CBF pair.
(ii) If we fix $\alpha=\alpha_1$ (Figure~\ref{fig: fully actuated cases}(a),(b)), the choice of $h$ significantly affects the region of attraction, whereas if we fix $\alpha=\alpha_2$ (Figure~\ref{fig: fully actuated cases}(c),(d)), the region of attraction remains almost the same for different $h$.
 We hypothesize that this is because the safety filter becomes inactive over a large region when the slope of $\alpha$ is large enough.
Therefore, the filter can only be active at points close to $\partial\mathcal{S}$. However, by Proposition~\ref{prop:jacobian-choice-cbf-safety-filters}, the spectrum of the Jacobian evaluated at the undesirable equilibria remains unchanged with different $h$ in the same equivalence class (like $h_1$ and $h_2$). 
All of this results in similar global dynamical behavior for different $h$ when the slope of $\alpha$ is sufficiently large.

\section{Conclusions}
\vspace{-0.3cm}
We have studied optimization-based control strategies for control-affine systems and investigated how the choice of the CBF impacts (desirable and undesirable) equilibria and the dynamical behavior of the resulting closed-loop system. We have shown that CBF-based constraints do not affect the number, location, and local stability properties of the equilibria in the interior of the safe set. We have also shown that undesirable equilibria only appear on the boundary of the safe set, and that their number and location do not depend on the choice of the CBF. For the specific case of safety filters, we have shown that the stability properties of the closed-loop system are the same for all CBF pairs for which the corresponding CBFs are in the same equivalence class. In future work, we plan to extend the stability results to other CBF-based control designs and explore the interplay between the choice of CBF pair and the region of attraction of equilibria.
\vspace{-0.3cm}

{
\small
\bibliographystyle{plainnat}        
\bibliography{alias,Main,Main-add,references}
}





\end{document}